# Ferromagnetism and insulating behavior with a logarithmic temperature dependence of resistivity in Pb$_{10-x}$Cu$_x$(PO$_4$)$_6$O

Yuhang Zhang, Cong Liu, Xiyu Zhu$^{*}$, and Hai-Hu Wen$^{*}$

*Center for Superconducting Physics and Materials, National Laboratory of Solid State Microstructures and Department of Physics, National Collaborative Innovation Center of Advanced Microstructures, Nanjing University, Nanjing 210093, China*

Recent claim of discovering above-room-temperature superconductivity ($T_c \approx 400$ K) at ambient pressure in copper doped apatite Pb$_{10}$(PO$_4$)$_6$O has stimulated world-wide enthusiasm and impulse of motion. A lot of follow-up works have been carried out with controversial conclusions. To check whether superconductivity is really present or absent in the material, we need samples which should have a rather pure phase of Pb$_{10-x}$Cu$_x$(PO$_4$)$_6$O. Here we report the characterization results from the Pb$_{10-x}$Cu$_x$(PO$_4$)$_6$O with a fraction of about 97 wt.% inferred from the fitting to the X-ray diffraction pattern, if assuming Cu$_2$S as the only impurity. The resistivity measurements show that it is a semiconductor characterized roughly by a ln(1/$T$) temperature dependence in wide temperature region without trace of superconductivity. Magnetization measurements show that it has a general ferromagnetic signal with a weak superparamagnetic background. Many grains of the sample show clear interactions with a NbFeB magnet. The detected Cu concentration is much lower than the expected nominal one. Our results show the absence of metallicity and superconductivity in Pb$_{10-x}$Cu$_x$(PO$_4$)$_6$O at ambient pressure, and suggest the presence of strong correlation effect.

superconductivity, Pb$_{10}$(PO$_4$)$_6$O, ferromagnetic, correlation effect

## 1  Introduction

Room temperature superconductivity has been regarded as the holy grail of condensed matter physics. Since the discovery of superconductivity by Onnes in 1911, tremendous efforts have been dedicated towards this goal, which has led to the discovery of two families of unconventional high-temperature superconductors at ambient pressure, namely, cuprates and iron-based superconductors [1,2]. Ashcroft [3,4] predicted that metallic hydrogen and hydrogen-rich materials under high pressures may be important platforms for exploring room-temperature superconductivity. In the last several years, under ultra-high pressures, the high temperature superconductivity in H$_3$S, LaH$_{10}$, CaH$_6$, and so on has been discovered [5-8]. However, the quest for room-temperature superconductivity did not succeed yet. The claimed room-temperature superconductors, like C-H-S system and N-doped lutetium hydride, were widely questioned and challenged [9-14]. Up to now, no material has been recognized as a room temperature superconductor at ambient pressure.

Very recently, a Cu doped Pb based apatite Pb$_{10-x}$Cu$_x$-(PO$_4$)$_6$O was claimed to be the first above-room-temperature superconductor at ambient pressure by Lee et al. [15,16], which attracted great attention in the community. Many groups have tried to replicate or disprove this discovery [17-25]. Few teams claimed to see the superconducting-like signs. For example, Wu et al. [17] reported the diamagnetic transition above 300 K and magnetically levitation with large levitated angle. A very small resistance at 110 K was found occasionally and was regarded as the "zero resistance" state by another group, but without the signature of Meissner effect [18]. However, the explicit evidence for showing superconductivity in this system is still lacking. Furthermore, Liu et al. [19] reported the semiconducting transport and absence of levitation in $Pb_{10-x}Cu_x(PO_4)_6O$. The half levitation was also observed by another group and attributed to the soft ferromagnetism [20]. Meanwhile, the sudden drop around 385 K in resistivity for $Pb_{10-x}Cu_x(PO_4)_6O$ was argued to be the first order structural phase transition of the $Cu_2S$ phase [21]. Some other groups also presented their data showing the absence of room temperature superconductivity in $Pb_{10-x}Cu_x(PO_4)_6O$ [22-25]. Despite the failed confirmation of room temperature superconductivity, interestingly, this material has attracted enormous attention from the theoretical point of view [26-30]. Some groups emphasized the importance of the flat bands around the Fermi level and argued that this might lead to the possibility of room-temperature superconductivity [26,27]. Other groups have made assessment about the correlation effect and suggest that the system may be a Mott or charge transfer insulator [28,29]. And another group demonstrates an interesting regime near the metal insulator transition driven by the charge transfer mechanism [30]. In addition, the experimental efforts for recognizing the magnetic properties of the system have led to quite diverse results [20,22,23]. It is highly desired to investigate the correlation effect and the properties of the intrinsic magnetic state.

*Corresponding authors (Xiyu Zhu, email: zhuxiyu@nju.edu.cn; Hai-Hu Wen, email: hhwen@nju.edu.cn)

In this manuscript, we successfully synthesized the $Pb_{10-x}Cu_x(PO_4)_6O$ samples with a purity higher than 97 wt.% by improving the reaction conditions. The sample exhibits a semiconducting behavior in the whole temperature region. A moderate ferromagnetic feature is detected, and some grains of the sample show clear interactions with a magnet. Our results also exclude the room-temperature superconductivity in present samples.

## 2 Experimental details

The samples of $Pb_{10-x}Cu_x(PO_4)_6O$ were synthesized by the solid-state reaction method as used in refs. [15,16]. Firstly, we obtained Lanarkite by sintering $PbSO_4$ powder (Meryer 99.99%) and PbO powder (Amethyst 99.99%) with the ratio of 1:1 at 725°C in air. The $Cu_3P$ was also obtained by the high temperature sintering with the mixed powders of Cu (Meryer 99.99%) and P (Aladdin 99.999%) in the ratio of 3:1. Then, the resultant Lanarkite and $Cu_3P$ were weighed in stoichiometric proportions (1:1), grounded, pressed into a pellet, sealed in an exhausted quartz tube, and then heated to 925°C and stayed for 10 h. It is found that the resultant sample contains large portion of $Pb_{10-x}Cu_x(PO_4)_6O$ with a dark color by this process, but also has impurity phases like $Pb_4(PO_4)_2(SO_4)$ and CuO. To inhibit the production of these impurity phases, a little bit of more phosphorus (about 0.13 molar ratio) was added in the reaction of the last step. Finally, we obtained the polycrystalline samples with the major phase of $Pb_{10-x}Cu_x(PO_4)_6O$ and the estimated concentration is higher than 97 wt.% as inferred from the refinement to the X-ray diffraction (XRD) pattern.

The crystal structure was evaluated via XRD (Bruker, D8 Advance diffractometer) with Cu Kα radiation ($\lambda$ = 1.5418 Å). The diffraction intensity was obtained over the $2\theta$ angle ranging from 10° to 90°. The Rietveld refinements [31] were conducted with TOPAS 4.2 software [32]. The dc magnetization measurements were performed with a SQUID-VSM-7 T (Quantum Design). A standard four-probe method was used to measure the electrical resistivity at ambient pressure with the physical property measurement system (PPMS, Quantum Design).

## 3 Results and discussion

### 3.1 Sample characterization

As depicted in the inset of Figure 1(a), the apatite $Pb_{10}$-$(PO_4)_6O$ phase has a hexagonal structure with the space group of $P63/m$. The XRD pattern with the Rietveld refinement is shown in Figure 1(a), and the scanning electron micrograph (SEM) images with energy dispersive spectroscopy (EDS) are shown in Figure 1(b). These data were measured to check the purity of our samples. From the refinements, the calculated profile agrees with our experimental data quite well. The pink vertical lines mark the indices belonging to the $Pb_{10-x}Cu_x(PO_4)_6O$ phase with the converted lattice parameters ($a$ = 9.854 Å, $c$ = 7.421 Å), while the phase indices marked with purple vertical lines are attributed to the $Cu_2S$ phase. One can see that, the refinement yields a content of 97.91 wt.% for the $Pb_{10-x}Cu_x(PO_4)_6O$ phase, and a very small amount of $Cu_2S$ with the ratio of 2.09 wt.% was also found, if we assume $Cu_2S$ as the only impurity. However, if some lead containing impurity phases exist, the estimated volume of $Pb_{10-x}Cu_x(PO_4)_6O$ would be higher. The nice fitting to the XRD data tells $Pb_{10-x}Cu_x$-$(PO_4)_6O$ is the major phase in our sample.

A comparison of the XRD data between our sample and that from Lee et al. [15,16] is given in Figure 2. One can see that our sample contains much less impurity phase of $Cu_2S$. We used the X-ray energy dispersive spectrum to analyze the compositions of the samples. As revealed by the table in Figure 1(b), for the ten spots investigated, the Cu content has some scattering; the mean value of the ratio of the two elements is roughly $Pb:Cu \approx 87:1$, which means that only a little copper was successfully substituted to the Pb sites. It seems to be a challenge to substitute more Cu for Pb at this moment.

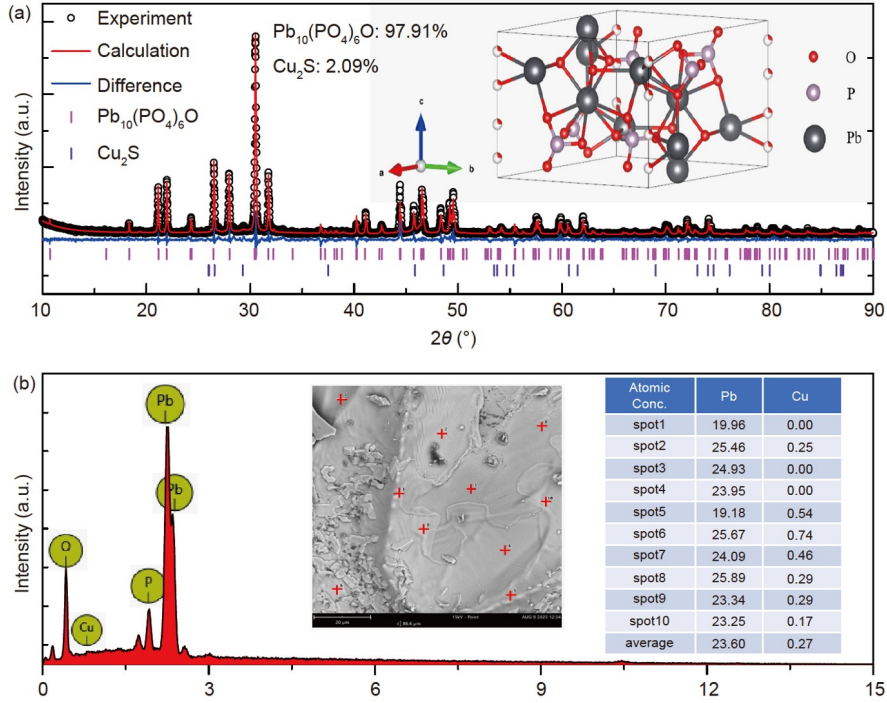

**Figure 1** (Color online) Structure and compositional analysis of $Pb_{10-x}Cu_x(PO_4)_6O$. (a) The XRD pattern for the polycrystalline sample $Pb_{10-x}Cu_x(PO_4)_6O$ with the Rietveld refinement. The pink vertical lines mark the indices belonging to the $Pb_{10-x}Cu_x(PO_4)_6O$ phase, and the purple vertical lines show the indices for the $Cu_2S$ phase. (b) A typical energy dispersive spectrum that shows the compositional ratio of the measured spots. The average molar ratio of lead and copper is $Pb:Cu \approx 87:1$. The inset (left) shows an image through a scanning electron microscope, and the inset (right) shows a table for the atomic concentration of Pb and Cu.

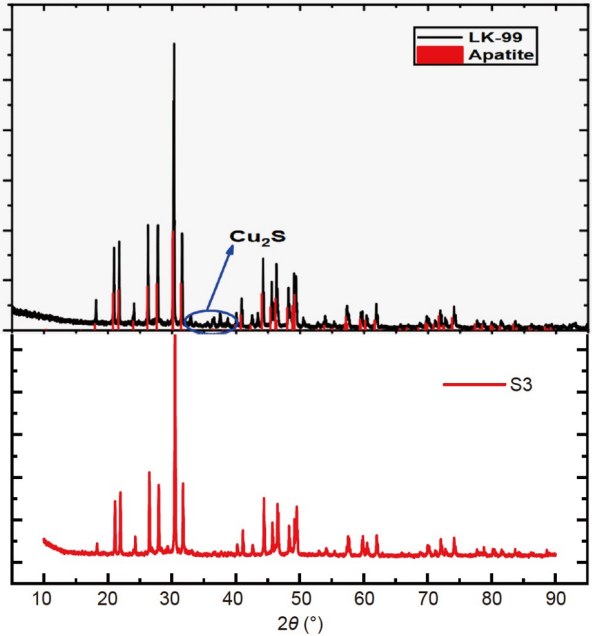

**Figure 2** (Color online) Comparison of XRD data between our sample (bottom) and that from Lee et al. [15,16] (upper). It shows that our sample contains less $Cu_2S$ impurity phase.

This is reasonable because the atomic radii of $Pb^{2+}$ and $Cu^{2+}$ are very different and it is very difficult to dope Cu to the Pb sites. Meanwhile, as reported by another group, the calculations show that Cu substitution is highly thermodynamically disfavored [33]. Our EDS data are consistent with this result.

### 3.2 Magnetic and electrical transport properties

Figure 3 shows the electric transport properties of the $Pb_{10-x}Cu_x(PO_4)_6O$ sample and the fitting results with different models. Figure 3(a) shows the temperature dependence of resistivity of two samples in different detecting channels, it is clear that resistivity increases with lowering temperature in the whole region, exhibiting a behavior of semiconductor or weak insulator. The inset of Figure 3(a) shows an image of two samples with four electrodes. We try to fit the temperature dependent resistivity with the formulas corresponding to different models, including the thermal activation model suitable for a band insulator ($\ln\rho \propto 1/T$), 3D variable-range-hopping (VRH) model [34,35] ($\ln\rho \propto T^{-1/4}$), and the model for correlation effect [36] ($\rho \propto \ln(1/T)$). The upper-right inset in Figure 3(b) was illustrated for the model of 3D VRH, while the lower-left inset illustrates the way of plotting data for the thermal activation model for a band insulator. If these models apply to the data here, they should show up as straight lines. However, it is clear that these models cannot capture the basic features of the temperature dependent resistivity; if we plot the data with $\rho$ versus ln(1/$T$), it shows roughly as a linear line in a wide temperature region. A small hump is observed in high temperature region, which may be induced by the small amount of Cu$_2$S, and in the low temperature region, a slight deviation is also observed, but the general behavior is roughly close to a linear relation. We have also fitted the resistivity data between 10-40 K, it is found that the thermal activation model suited for the band insulator still exhibits a large deviation. However, the data for the VRH and $\rho$-ln($T$) model show rough linear behavior; because the VRH model shows clear deviation in the high temperature region, we thus believe our data better fit to the $\rho$-ln($T$) model. Usually, this kind of logarithmic temperature dependence can be explained by the Kondo or correlation effect; however, Kondo effect would happen only in low temperature region due to the formation of Kondo singlet. Interestingly, in some cuprate samples, like underdoped Bi$_2$Sr$_{2-x}$La$_x$CuO$_{6+\delta}$ and La$_{2-x}$Sr$_x$CuO$_4$, the logarithmic behavior ($\rho \propto$ ln(1/$T$)) was also observed [37,38], which was argued to be a key point related to the strong correlation effect. Up to now, the origin for this behavior remains still elusive, but a general argument of linking this behavior with strong correlation effect seems conceivable. From the analysis mentioned above, the models with a band insulator and the 3D VRH cannot explain the behavior of resistivity data, but the transport data strongly suggest the involvement of strong correlation effect. We must mention that at this moment we cannot exclude the influence of the scattering by the grain boundaries, but this may not change the general argument and needs to be further checked with single crystal samples. From the above analysis, we would conclude that the semiconducting behavior in present sample is governed by the logarithmic temperature dependence which is intimately induced by the strong correlation effect, as proposed by several theoretical studies [28-30]. We must emphasize that the Cu doping level is inhomogeneous in different grains, although the general doping level is rather low. Thus, the logarithmic temperature dependence of resistivity may be an average effect between different grains. However, the fitting results of other models seriously deviate from the linear one. We believe that the $\rho \propto$ ln(1/$T$) behavior is closer to the intrinsic transport property of the sample.

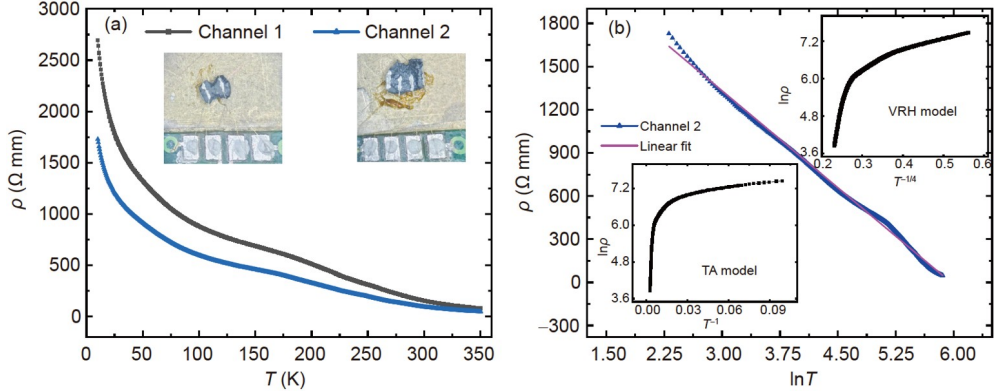

**Figure 3** (Color online) Electric transport properties of Pb$_{10-x}$Cu$_x$(PO$_4$)$_6$O. (a) Temperature dependence of resistivity for the Pb$_{10-x}$Cu$_x$(PO$_4$)$_6$O samples in the temperature range from 10 to 350 K at zero magnetic field and ambient pressure. (b) Temperature dependence of resistivity in the form of $\rho$ versus ln($T$) for the Pb$_{10-x}$Cu$_x$(PO$_4$)$_6$O sample from 10 to 350 K with corresponding linear fit (pink curve). The upper right inset shows the curve of ln$\rho$ versus $T^{-1/4}$ for the VRH model, and the lower left inset shows the curve of ln$\rho$ versus $T^{-1}$ expected for a band insulator.

To explore the magnetic properties of Pb$_{10-x}$Cu$_x$(PO$_4$)$_6$O, we measured the dc magnetization $M(T)$ and magnetic hysteresis loops (MHL) by the SQUID-VSM magnetometer, the data are shown in Figure 4. The curves of temperature dependent magnetic moment at 10 Oe and 10 kOe detected in the zero-field-cooling (ZFC) and field-cooling process (FC) mode are shown in Figure 4(a). The magnetic moment shows a positive value in the whole temperature region, but without showing a Curie transition up to 350 K. The MHLs acquired from our sample at different temperatures are shown in Figure 4(b). One can see that the MHLs exhibit a clear but moderate ferromagnetic signal. As the temperature decreases, the hysteresis becomes more pronounced. Besides the hysteresis, a superparamagnetic signal exists when the FM state comes to saturate. The saturated magnetic moment seems to be small or moderate, which suggests that the ferromagnetic phase may be induced by the diluted dopants or very small magnetic moments. We are also concerned about the absence of ferromagnetism and half-levitation effect, but the presence of diamagnetism in another report [22]. Our results are partially consistent with the results in ref. [20], which shows the ferromagnetism and successfully explains the half-levitation effect. We must point out that during the synthesis process, a slight excess of phosphorus is added to inhibit sulfate residues and obtain a purer sample. This may lead to a little bit difference of magnetization between our sample and that from other groups [20-22]. There are several reasons to explain the different magnetic properties. Firstly, the variation of copper doping may lead to this discrepancy.

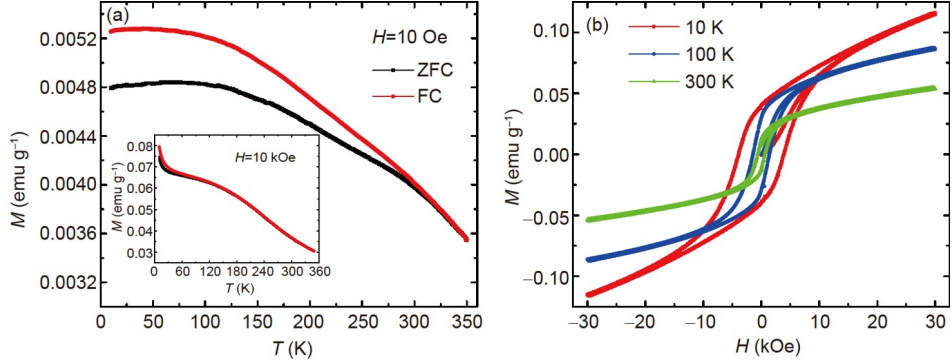

**Figure 4** (Color online) Magnetic properties for $Pb_{10-x}Cu_x(PO_4)_6O$. (a) Temperature dependence of magnetic moment measured on one sample of $Pb_{10-x}Cu_x(PO_4)_6O$ (mass = 81.50 mg) at 10 Oe and 10 kOe (inset). (b) Magnetic hysteresis loops measured at different temperatures (10, 100, and 300 K), showing a moderate ferromagnetic behavior together with a superparamagnetic feature.

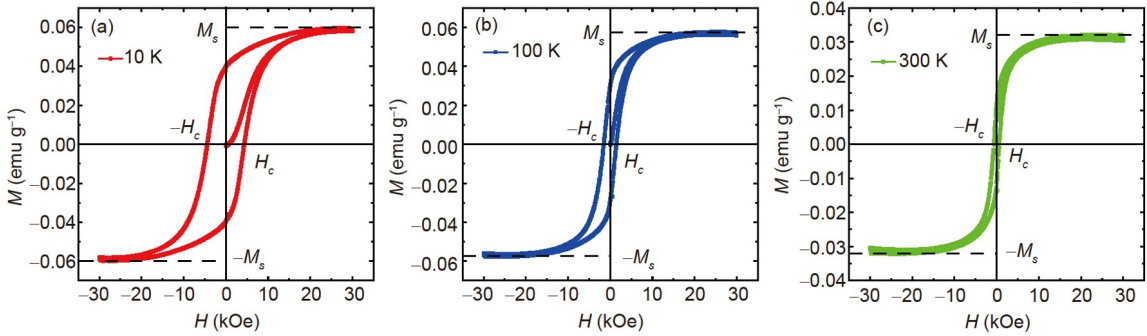

**Figure 5** (Color online) MHLs at 10 (a), 100 (b), and 300 K (c) with the superparamagnetic signal subtracted. Now the MHL data contain only the signal of ferromagnetism. The coercivity fields $H_c$ and the saturated magnetic moments $M_s$ are marked on the curves.

As indicated by the EDS data, the copper content varies in different regions in our sample. Secondly, in the crystal structure, there is an uncertainty for the oxygen atom at 4e position: on one hand, the oxygen atom occupancy at this position is 1/4, and the variation of the occupancy may affect the magnetic properties; on the other hand, according to the report from another group [33], the oxygen atom at this position may combine with $H_2O$ to form $Pb_{10}(PO_4)_6(OH)_2$. Theoretical calculations show that, whether it is $Pb_{10}(PO_4)_6O$ or $Pb_{10}(PO_4)_6(OH)_2$, the Cu doping tends to induce a ferromagnetic phase. More works are needed to resolve this discrepancy of magnetic properties.

We fit the superparamagnetic term as a linear line crossing the origin and subtract this background, the data for the pure ferromagnetic signal are shown in Figure 5. Now we can see the standard MHLs for ferromagnetism, and they can yield the saturated magnetic moments $M_s$, which are about 0.0599 emu $g^{-1}$ (1 emu $g^{-1}$ = 1 A $m^2$ $kg^{-1}$) at 10 K, 0.0575 emu $g^{-1}$ at 100 K and 0.0321 emu $g^{-1}$ at 300 K, respectively. Meanwhile, the coercivity fields $H_c$ at the temperatures of 10, 100, and 300 K are determined to be 4376, 1492, and 665 Oe (1 Oe = 79.5775 A/m), respectively. Thus, we regard the ferromagnetic moment in the system as moderate. We also use a NbFeB magnet to approach the powders of the sample, and it is found that many small grains show the clear interactions with the magnet by moving when we shake the magnet. A video for a piece of sample showing the so-called half-levitation effect is presented in Supporting Information (Video S1). It is clear that the sample can be lifted vertically, demonstrating an interaction. We believe this half-levitation is mainly induced by the ferromagnetic interaction, consistent with a recent report and judgement [20]. We become aware that the samples synthesized by another group [22] show the absence of the ferromagnetic signal, but only a linear diamagnetic signal to a high magnetic field was observed, although the XRD of that sample can also be fitted nicely by the simulation to the $Pb_{10}(PO_4)_6O$. The ferromagnetic signal and the logarithmic temperature dependence of resistivity in our samples can be repeated. However, it should be pointed out that the ferromagnetic signal and the logarithmic temperature dependence of resistivity may also rely on the oxygen content, hydration levels or different doping levels of Cu, but unfortunately, we cannot find related data of Cu composition from EDS analysis in the exposed works so far. This discrepancy also suggests that the $Pb_{10-x}Cu_x(PO_4)_6O$ system possesses by itself many variable changes upon varying the Cu concentration, different oxygen content, or annealing and ageing process. Our control experiments on the samples with black color reveal a moderate ferromagnetism and insulating behavior. Further efforts are needed to clarify the distinctions of magnetism on the samples from different groups.

## 4 Conclusions

We have successfully synthesized the sample $Pb_{10-x}Cu_x$-$(PO_4)_6O$ with estimated content reaching 97 wt.%, if taking $Cu_2S$ as the only impurity. The resistivity shows a semiconducting behavior and a logarithmic temperature dependence ($\rho \propto \ln(1/T)$) in a wide temperature region, which suggests a strong correlation effect in the system. The dc magnetization data, $M(T)$ and MHLs exhibit the ferromagnetic behavior with a moderate saturated magnetic moment in $Pb_{10-x}Cu_x(PO_4)_6O$. The EDS results show a low concentration of Cu in the sample, indicating the difficulty to dope Cu to the Pb sites. From both resistivity and magnetization, no signature of superconductivity was found.

*This work was supported by the National Key R&D Program of China (Grant No. 2022YFA1403201), the National Natural Science Foundation of China (Grant Nos. 12061131001, 52072170, 12204231, and 11927809), and the Strategic Priority Research Program (B) of Chinese Academy of Sciences (Grant No. XDB25000000).*

**Conflict of interest** *The authors declare that they have no conflict of interest.*